\documentclass[aps, prl, twocolumn, superscriptaddress, notitlepage,floatfix]{revtex4-2}
\usepackage{float}
\usepackage{bbm}
\usepackage{epsfig}
\usepackage{epstopdf}
\usepackage{graphicx}
\usepackage{amsmath,amssymb}
\usepackage{amsmath,bm}
\usepackage{amsmath,esint}
\usepackage{physics}
\usepackage{color}
\usepackage{hyperref}
\usepackage{url}
\usepackage{lineno,blindtext}
\usepackage[caption=false]{subfig}
\usepackage{siunitx, booktabs}
\usepackage{diagbox, eqparbox, hhline}
\usepackage{soul}
\usepackage{xcolor}
\usepackage[normalem]{ulem}
\newcolumntype{P}[1]{>{\centering\arraybackslash}p{#1}}
\begin{document}

\title{Collective Coordinate Dynamics of Antiferromagnet: Gravity by Quantum Metric}

\author{Kuangyin Deng}
\email{kuangyid@ucr.edu}
\affiliation{Department of Electrical and Computer Engineering,  University of California, Riverside, California 92521, USA}
\author{Ran Cheng}
\email{rancheng@ucr.edu}
\affiliation{Department of Electrical and Computer Engineering,  University of California, Riverside, California 92521, USA}
\affiliation{Department of Physics and Astronomy, University of California, Riverside, California 92521, USA}
\affiliation{Department of Materials Science and Engineering, University of California, Riverside, California 92521, USA}

\begin{abstract}
The collective coordinate dynamics of antiferromagnetic and ferrimagnetic textures emerges as a geodesic equation governed by the quantum geometry of the CP$^1$ state associated with the Néel vector, rendering an effective Riemannian gravity and a Lorentz force in the texture motion. This unveils a physical scenario fundamentally distinct from the Thiele equation widely used for ferromagnetic textures, which is inertialess without any gravity analogy. Applying our framework to an optically driven ultrafast antiferromagnetic domain wall, we can explain the helicity-dependent asymmetric motion observed experimentally and predict a vibrantly breathing wall width.
\end{abstract}

\maketitle 

In studying the dynamics of magnetic textures, collective coordinates (CCs) provide a highly effective description that recasts the coupled dynamics of massive number of spins into just a few characteristic degrees of freedom, such as the position and width of a topological defect~\cite{Manton2004topological}. Thiele's equation, being the most well-known CC model among its kind, captures the collective behavior of ferromagnetic textures as the interplay among gyrotropic, dissipative, and external forces~\cite{Thiele1973,kim2023generalizing,weibenhofer21skyrmion}. Owing to its remarkable predictive power, Thiele's equation has become a central framework for studying various ferromagnetic textures such as domain walls and skyrmions~\cite{Tretiakov2008,Zang2011,Lin2013,Nagaosa2013topological}.

Recent explorations have generalized the framework of CCs into antiferromagnetic (AFM) textures~\cite{Tveten2013,Velkov2016,Cheng2014PRB,Shiino2016,Tveten2016intrinsic,Gomonay2018antiferromagnetic,wu2022current}, where the Néel-vector dynamics is intrinsically second order in time derivative (\textit{i.e.}, entails inertial), resulting in a Newton-like equation without the gyrotropic term. Existing investigations along this line, however, are predominantly limited to slow motions, which could be significantly broken when an AFM texture moves at ultrafast speeds approaching the magnon velocity~\cite{Seyler2025high}. More seriously, we find an improper approximation with critical missing in existing theories, which, as will be elaborated later, could be invalid even for a steady motion.

In this Letter, we construct a general CC dynamics free of the above problems and naturally incorporates both AFM and ferrimagnetic textures in a unified fashion. To highlight the gist of our findings, let us first illustrate the central idea in plain language. Adopting a set of CCs denoted by $\xi^\mu$, we will derive the texture's evolution with time as a hypothetical particle moving in a curved space (\textit{i.e.}, the CC manifold). The effective dynamics assumes the elegant form of a geodesic equation involving both gravity and non-gravitational forces:
\begin{align}
\ddot{\xi}^\rho+\Gamma^{\rho}_{\mu\nu}\dot{\xi}^\mu\dot{\xi}^\nu=\mathcal{F}^\rho-\alpha\omega_0M\dot{\xi}^\rho+\frac12\omega_0\delta M\Omega^{\rho}_{\,\mu}\dot{\xi}^\mu,
\label{eq:ddn-final}
\end{align}
where $\alpha$ is the Gilbert damping constant, $\omega_0$ is the exchange frequency, and $\delta M$ is the imbalance of sublattice magnetization so that $\delta M\rightarrow0$ refers to the AFM limit. According to Eq.~\eqref{eq:ddn-final}, the texture motion reflected by the CC dynamics is induced by four types of forces: (1) the curved CC manifold provides a background gravity through the affine connection $\Gamma^{\rho}_{\mu\nu}$, (2) the driving force $\mathcal{F}^\rho$ originates from both internal magnetic interactions and external stimuli, (3) $\alpha\omega_0M\dot{\xi}^\rho$ plays the role of a friction, and (4) the Berry curvature term $\Omega^{\rho}_{\,\mu}\dot{\xi}^\mu$ exerts a Lorentz force that vanishes in the AFM limit $\delta M\rightarrow0$. Among these ingredients, $\Gamma^{\rho}_{\mu\nu}$ and $\Omega^{\rho}_{\,\mu}$ are directly related to the quantum geometric tensor (QGT)~\cite{ProvostVallee1980,Kolodrubetz2017geometry}, $\mathcal{Q}_{\mu\nu}=\mathcal{G}_{\mu\nu}-\mathrm{i}\Omega_{\mu\nu}/2$, such that $\Omega^{\rho}_{\,\mu}=\mathcal{G}^{\rho\beta}\Omega_{\beta\mu}$ with $\mathcal{G}^{\mu\nu}$ being the inverse matrix of the quantum metric $\mathcal{G}_{\mu\nu}$. For an instructive comparison, Thiele's equation in terms of the same notations read:
\begin{align}
 \omega_0M\left(\Omega_{\mu\nu}-2\alpha \mathcal{G}_{\mu\nu}\right)\dot{\xi}^\nu+2\mathcal{F}_\mu=0,
 \label{eq:Thiele}
\end{align}
in which $\Omega_{\mu\nu}$ and $\mathcal{G}_{\mu\nu}$ generate a gyroscopic force and a dissipative force, respectively, without any gravity.

While quantum metric and Berry curvature are juxtaposed under the QGT, their dynamical implications have not been recognized to the same level in literature. The last few decades have witnessed ubiquitous phenomena where Berry curvature behaves directly as an electromagnetic field by generating Lorentz-like forces. By contrast, quantum metric has rarely been associated directly to gravity, but rather, manifests indirectly through nonlinear responses~\cite{Gao2025quantum,Liu2025quantum,Ahn2020,Gao2023,Wang2023quantum,Das2023intrinsic}, instability~\cite{PeottaTorma2015,Julku2016,Hu2019geometric,Xie2020topology,Julku2021quantum}, electron localization~\cite{MarzariVanderbilt1997,Resta2011insulating,Onishi2024qw}, and so on. It is only recently that quantum metric is identified as an emergent gravity affecting the motion of Bloch electrons in the phase space~\cite{MomentumSpaceGravity,Mehraeen2025,Ren2026nonadiabatic,GravityLense2026}. Nevertheless, the dynamics of magnetic textures remains an uncharted territory in the quest for gravitational effects of quantum metric.

Timely filling this knowledge gap, our Eq.~\eqref{eq:ddn-final} reveals a genuine gravitational nature of the quantum metric associated with the AFM order parameter, which entails profound consequences in the texture motion. As a concrete example, we apply our theory to a recent experiment~\cite{Seyler2025high} and explain the observed asymmetric motion of an ultrafast AFM domain wall (DW) through the emerging gravity of quantum metric, resolving a crucial puzzle left by the phenomenological model therein. In addition, we predict that the DW motion is accompanied by a highly oscillating DW width, consistent with a hidden feature unpublished with the experiment~\cite{PC}.

\textit{Order parameter dynamics}---We consider a collinear magnet having two sublattices characterized by dimensionless local moments $\bm m_{1,2}(\bm{r},t)$ of fixed magnitudes $M_{1,2}$ (so $M_1=M_2$ defines the AFM limit), which varies slowly over space and time. Suppose that the system has a well-defined free energy $U$, and each sublattice $\bm{m}_i=\bm{m}_i(\bm{r},t)$ is locally subject to a damping-like spin torque $\bm\tau_i^{\rm D}$ (while field-like torques can be absorbed by $U$), the magnetic dynamics is described by the coupled Landau-Lifshitz-Gilbert equations
\begin{align}
\dot{\bm m}_i=&-\bm m_i\times \bm H_i^{\rm eff}
+\frac{\alpha}{M_i}\bm m_i\times\dot{\bm m}_i
+\bm\tau_i^{\rm D},
\label{eq:llg-main}
\end{align}
where the effective field $\bm{H}_i^{\rm eff}(\bm{r},t)=-(\gamma/\mathcal{M}_d)\delta U/\delta\bm{m}_i$ absorbs the gyromagnetic ratio $\gamma$, with $\mathcal{M}_d$ the magnetization density on sublattice $i$.

By using the normalized vectors $\hat{\bm m}_i=\bm m_i/M_i$, we can treat compensated AFM ($M_1=M_2$) and ferrimagnetic ($M_1\neq M_2$) cases on an equal footing regardless of the ratio $M_1/M_2$~\cite{FiM2022Guo,wu2022current}. As such, the system is characterized by the local Néel order $\bm n=(\hat{\bm m}_1-\hat{\bm m}_2)/2$ and a small magnetic vector $\bm m=(\hat{\bm m}_1+\hat{\bm m}_2)/2$ with $M=M_1+M_2$ and $\delta M=M_1-M_2$, satisfying $\bm{n}\cdot\bm{m}=\frac14(\hat{\bm m}_1^2-\hat{\bm m}_2^2)=0$. In the exchange regime, $|\bm n|\approx 1$ and $|\bm m|\ll 1$, and $\bm m$ can be customarily eliminated from Eq.~\eqref{eq:llg-main}. Keeping the leading terms under the approximation $\lambda_c\gg a$, where $\lambda_c$ represents the characteristic length of the texture and $a$ is the lattice constant, yields the effective dynamics of the N\'eel order~\cite{SM}
\begin{align}
\ddot{\bm n}+\omega_0\delta M\bm n\times \dot{\bm n}+\alpha\omega_0M\dot{\bm n}
=\bm q(\bm n,\dot{\bm n}),
\label{eq:n-eom-main}
\end{align}
where the right-hand side is a generalized force, and $\omega_0$ is the exchange angular frequency---the largest frequency scale in the system. In the AFM limit that $\delta M\rightarrow0$, the $\bm{n}\times\dot{\bm{n}}$ term vanishes, retrieving what is commonly known as the nonlinear $\sigma$-model~\cite{Haldane1983}.

Our formalism is constructed in a generic setting that does not reply on any specific conditions. To better visualize the physical picture, especially the form of $\bm{q}(\bm{n},\dot{\bm{n}})$, we now consider a prototype 2D AFM texture. Scaling all parameters into areal densities, we include in the free energy the antiferromagnetic exchange coupling $J_0>0$ (with the coordination number absorbed), the easy-axis anisotropy $K_0>0$, a Zeeman field $H_0$ applied in the $z$ direction, and an interfacial Dzyaloshinskii-Moriya interaction due to mirro-symmetry breaking
\begin{align}
 \epsilon_i^{\rm DM}=D_0\left[m_{iz}(\nabla_\perp \cdot \bm m_i)-(\bm m_i \cdot \nabla_\perp)m_{iz}\right] \nonumber
\end{align}
with $\nabla_{\perp}=(\partial_x,\partial_y,0)$. Additional interactions can be easily incorporated so long as $J_0$ remains dominant. In the continuum limit, the free energy is~\cite{SM}
\begin{align}
U=&\int d^2r\Bigg\{J_0\bm{m}_1\cdot \bm{m}_2+A_0\bm{m}_1\cdot\Diamond \bm{m}_2-P_0\bm{m}_1\cdot\partial_+\bm{m}_2 \nonumber\\
&\left.+\sum_{i=1,2}\left[\epsilon_i^{\rm{DM}}-\frac12K_0(\bm{m}_i\cdot\hat{\bm{z}})^2-H_0\hat{\bm z}\cdot\bm m_i\right]\right\},\label{eq:free-energy-main}
\end{align}
where $A_0=J_0a^2/4$, $P_0=J_0a/2$, and $\partial_+\equiv \partial_x+\partial_y$ and $\Diamond \equiv\partial_x^2+\partial_y^2+\partial_x\partial_y$. The parity-breaking term, proportional to $P_0$, arises naturally from the coarse-grained staggered texture. In this 2D example, the effective field acting on $\bm{m}_1$ is
\begin{align}
 \bm H^{\rm eff}_1=&\omega_K m_{1z}\hat{\bm z}+\omega_H\hat{\bm z}
+2v_{\rm DM}\left[\nabla_\perp m_{1z}-(\nabla_\perp \cdot \bm m_1)\hat{\bm z}\right]\nonumber\\
&-\omega_0\bm m_2+P\partial_+\bm m_2-A\Diamond \bm m_2,
\label{eq:heff-main}
\end{align}
and its counterpart $\bm H^{\rm{eff}}_2$ is obtained by exchanging $\bm{m}_1\leftrightarrow\bm{m}_2$ while flipping the sign of the $P$ term. In Eq.~\eqref{eq:heff-main}, every term has been scaled into the dimension of angular frequency, and their coefficients are:
\begin{align}
 (\omega_K,\omega_H,v_{\rm DM},\omega_0,P,A)=\frac{\gamma}{\mathcal{M}_0t_0}(K_0,H_0,D_0,J_0,P_0,A_0) \nonumber
\end{align}
where $t_0$ is the sample thickness and $\mathcal M_0$ serves as a reference parameter making $\mathcal{M}_d=\mathcal{M}_0t_0$ the areal magnetization density of sublattice $i$. Suppose that the damping-like torque arises from a uniform non-equilibrium spin polarization $\hat{\bm{p}}$, \textit{i.e.}, $\bm\tau_i^{\rm D}=(\omega_s/M_i)\bm{m}_i\times(\hat{\bm{p}}\times\bm{m}_i)$, the generalized force in Eq.~\eqref{eq:n-eom-main} is obtained as
\begin{align}
\bm{q}&=\frac{\omega_0\omega_K}2M^2 n_z\hat{\bm z}-\omega_0\omega_H\delta M(n_z\bm{n}-\hat{\bm z})+\omega_H\hat{\bm z}\times\dot{\bm{n}} \nonumber\\
&\quad+\omega_0v_{\rm{DM}}M^2\left[\nabla_{\perp}n_z-(\nabla_{\perp}\cdot\bm{n})\hat{\bm z}\right]-\omega_0\omega_sM\hat{\bm p}\times\bm{n}\nonumber\\
&\quad+\frac{\omega_0}4AM^2\nabla^2\bm{n}-\frac{\omega_0}2PM\delta M\partial_+\bm{n},
\label{eq:force-main}
\end{align}
where the mixed derivative $\partial_x\partial_y$ originating from $\Diamond$ got canceled out. We emphasize that the above 2D example is only a demonstration, whereas our formalism is general and free from system-specific restrictions.

\textit{QGT and CC dynamics}---Similar to the case of ferromagnetic textures, here the order parameter $\bm{n}$ depends on time through a set of CCs denoted by $\xi(t)=\{\xi^\mu(t)\}$, which could refer to any low-energy degree of freedom such as the position, phase and width of a topological defect. By the chain rule, $\dot{\bm n}=\partial_\mu\bm n\dot{\xi}^\mu$ and
\begin{align}
 \ddot{\bm n}=\partial_\mu\bm n\ddot{\xi}^\mu+\partial_\mu\partial_\nu\bm{n}\dot{\xi}^\mu\dot{\xi}^\nu,
 \label{eq:ndd}
\end{align}
where the second term had been discarded in previous studies for being quadratic in velocity~\cite{Tveten2013,Cheng2014PRB,Velkov2016}. However, comparing $\dot{\xi}^\mu\dot{\xi}^\nu$ directly against $\ddot{\xi}^\mu$ is not by itself a reliable criterion while their coefficients explicitly depend on $\xi^\mu$. For instance, a steady-state motion in the CC space has a vanishing acceleration $\ddot{\xi}^\mu\rightarrow0$, hence the first term in Eq.~\eqref{eq:ndd} vanishes, rendering the second term essential. Moreover, for ultrafast CC dynamics, $\partial_\mu\partial_\nu\bm{n}\dot{\xi}^\mu\dot{\xi}^\nu$ can even be comparable to the dominant exchange-force term proportional to $\nabla^2\bm{n}$ in Eq.~\eqref{eq:force-main}~\cite{footnote}.

At each real-space location $\bm r$, the field $\bm n[\bm r,\xi(t)]$ maps the CC manifold to a target space $S^2$. Since $S^2$ is diffeomorphic to $CP^1$ (1D complex projective space), this configuration can be represented by a normalized projective spinor $|z(\bm{r},\xi)\rangle$ determined up to a local $U(1)$ transformation, serving as the $CP^1$ representation of the nonlinear $\sigma$-model~\cite{Auerbach2012interacting,Banerjee1994quantum,Cheng2010equivalence}. Correspondingly, the QGT is locally defined as $Q_{\mu\nu}(\bm{r},\xi)=\langle\partial_\mu z|(1-|z\rangle\langle z|)|\partial_\nu z\rangle$ \cite{Gao2025quantum,Liu2025quantum}. With the gauge convention $|z(\theta,\phi)\rangle=(-e^{-\rm i\phi}\sin\frac{\theta}{2},\,\cos\frac{\theta}{2})^T$, the order parameter is related to the $CP^1$ field by the Hopf map $\langle z|\bm\sigma|z\rangle=-\bm n$ where $\bm{\sigma}=\{\sigma_x,\sigma_y,\sigma_z\}$ are the Pauli matrices. The QGT can be decomposed into $Q_{\mu\nu}=g_{\mu\nu}-{\rm i}\ell_{\mu\nu}/2$ where $g_{\mu\nu}(\bm{r},\xi)\equiv(1/4)\partial_\mu\bm{n}\cdot\partial_\nu\bm{n}$ and $\ell_{\mu\nu}(\bm{r},\xi)\equiv (1/2)\bm{n}\cdot(\partial_\mu\bm{n}\times\partial_\nu\bm{n})$ are the quantum metric and the Berry curvature, respectively, which transform covariantly under CC reparameterization~\cite{covariance}.

By integrating $Q_{\mu\nu}$ with respect to $\bm{r}$ over the entire system, we obtain the corresponding geometric quantities defined on the CC manifold as
\begin{subequations}
\label{eq:GLF-def}
\begin{align}
\mathcal{G}_{\mu\nu}(\xi)=&\frac{1}{A_{\rm s}}\int d^2r g_{\mu\nu}(\bm r,\xi), \label{eq:G}\\
\Omega_{\mu\nu}(\xi)=&\frac{1}{A_{\rm s}}\int d^2r \ell_{\mu\nu}(\bm r,\xi),
\end{align}
\end{subequations}
where $A_{\rm s}$ is the system area. Here $\mathcal{Q}_{\mu\nu}=\mathcal{G}_{\mu\nu}-\mathrm{i}\Omega_{\mu\nu}/2$ is the QGT defined on the CC manifold, characterizing the collective motions of the magnetic texture (all spins are involved). Besides the base manifold, another subtlety underlying Eqs.~\eqref{eq:GLF-def} pertains to the matrix invertibility. At a given location $\bm r$, the $\bm{n}$ field lives on a unit sphere, so the tangent vectors $\partial_\mu\bm n$ span at most a 2D tangent plane regardless of the dimension of $\xi^\mu$, hence $\mathrm{rank}\,g_{\mu\nu}(\bm r)\leq2$ everywhere. As a result, if more than two CCs exist, $g_{\mu\nu}$ becomes singular. On the other hand, each element of $\mathcal{G}_{\mu\nu}$ is a spacial overlap of two continuous functions in the whole domain $A_{\rm s}$. Therefore, $\mathcal{G}_{\mu\nu}(\xi)$ is invertible so long as the CCs are independent variables, even when $g_{\mu\nu}(\bm r)$ becomes locally singular.

Substituting $\bm n=\bm n[\bm r,\xi(t)]$ into Eq.~\eqref{eq:n-eom-main}, and taking the inner product with $\partial_\beta\bm n$ on both sides, we get
\begin{align}
g_{\beta\mu}\ddot{\xi}^\mu
&+\frac14\partial_\beta\bm n\cdot(\partial_\mu\partial_\nu\bm n)\dot{\xi}^\mu\dot{\xi}^\nu\nonumber\\
&-\frac12\omega_0\delta M\ell_{\beta\mu}\dot{\xi}^\mu+\alpha \omega_0Mg_{\beta\mu}\dot{\xi}^\mu=f_\beta,\label{eq:n-eom-main-1}
\end{align}
where $f_\beta\equiv\frac14\bm{q}\cdot\partial_\beta\bm n$. Integrating Eq.~\eqref{eq:n-eom-main-1} with respect to $\bm{r}$ over the area $A_{\rm s}$, we obtain
\begin{align}
&\mathcal{G}_{\beta\mu}\ddot{\xi}^\mu+\frac{1}{4A_{\rm s}}\int d^2r\partial_\beta\bm n\cdot(\partial_\mu\partial_\nu\bm n)\dot{\xi}^\mu\dot{\xi}^\nu\nonumber\\
&\qquad-\frac12\omega_0\delta M\Omega_{\beta\mu}\dot{\xi}^\mu
+\alpha\omega_0M\mathcal{G}_{\beta\mu}\dot{\xi}^\mu=\mathcal{F}_\beta,
\label{eq:n-eom-main-2}
\end{align}
where
\begin{align}
 \mathcal{F}_\beta(\xi)=&\frac{1}{A_{\rm s}}\int d^2r f_\beta(\bm r,\xi)
 \label{eq:genF}
\end{align}
is the generalized force acting on the CCs. The inverse of $\mathcal{G}_{\mu\nu}$ corresponds to the contra-variant matrix $\mathcal{G}^{\mu\nu}$ defined by $\mathcal{G}^{\mu\beta}\mathcal{G}_{\beta\nu}=\delta^\mu_{\,\nu}$. Correspondingly, the Christoffel symbols (affine connection) are
\begin{align}
\Gamma^\rho_{\mu\nu}\equiv&\frac{1}{2}\mathcal{G}^{\rho\beta}\big(\partial_\mu \mathcal{G}_{\nu\beta}+\partial_\nu \mathcal{G}_{\mu\beta}-\partial_\beta \mathcal{G}_{\mu\nu}\big)\nonumber\\
=&\frac{1}{4A_{\rm s}}\mathcal{G}^{\rho\beta}\int d^2r\partial_\beta\bm n\cdot(\partial_\mu\partial_\nu\bm n).
\label{eq:Gamma-collective}
\end{align}
Finally, raising indices with $\mathcal{G}^{\rho\beta}$ on both sides of Eq.~\eqref{eq:n-eom-main-2} and considering Eq.~\eqref{eq:Gamma-collective}, we arrive at the central result Eq.~\eqref{eq:ddn-final}---the geodesic equation governing the CC dynamics for AFM and ferrimagnetic textures.

In our formalism, the interactions between the AFM texture and various external stimuli may not always be incorporable by the free energy $U$, such as the damping-like spin torques, but in principle the generalized force $\bm{q}(\bm{n},\dot{\bm{n}})$ can include all kinds of interactions, so does $f_\mu(\bm{r},\xi)$ and $\mathcal{F}_\mu(\xi)$ appearing in Eq.~\eqref{eq:genF}. We also mention that $\mathcal{G}_{\mu\nu}$ furnishing the CC manifold is a Riemannian metric, unlike the pseudo-Riemannian metric of the real spacetime~\cite{covariance}.

\begin{figure}[t]
\centering
\includegraphics[width=\linewidth]{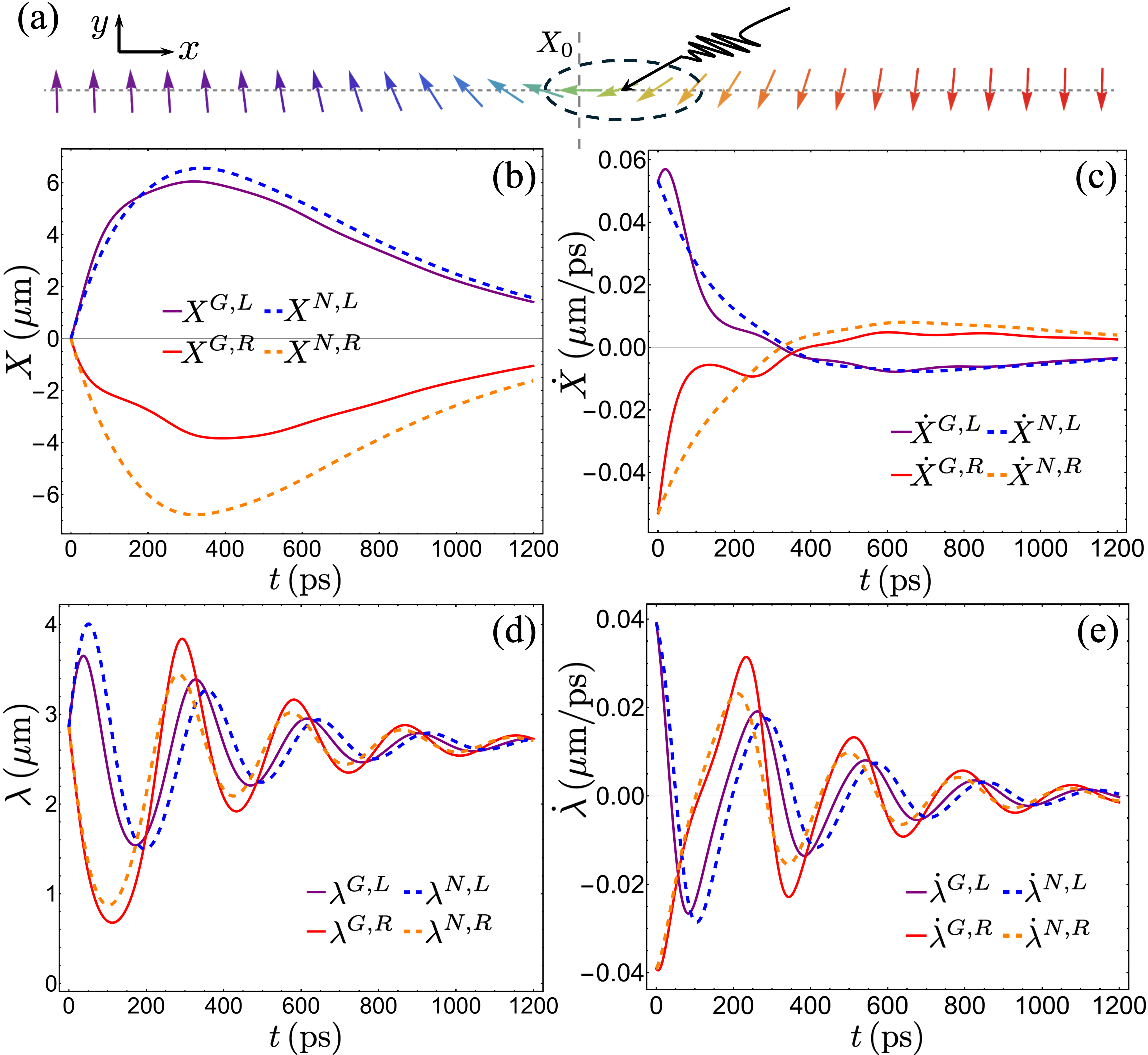}
\caption{Numerical results of the CC dynamics based on the optically driven AFM DW reported in Ref.~\cite{Seyler2025high}. \textbf{(a)} Schematic illustration of an AFM DW along the $x$ direction with winding number $w=-1/2$. An optical pump with Gaussian profile, centered at $x_0$, is marked by the dashed ellipse and is displaced from the initial DW center $X_0$. The colored arrows indicate the local Néel vector $\bm{n}$. \textbf{(b)} and \textbf{(c)}: DW center $X(t)$ and DW velocity $\dot{X}(t)$ as functions of time. In the super-indices, left (L) and right (R) refer to the helicity of the laser pulse, while (G) and (N) distinguishes results obtained from the full geodesic equation and that from the Newtonian approximation by dropping the connection term. The solid curves show a pronounced helicity-dependent asymmetry. \textbf{(d)} and \textbf{(e)}: DW width $\lambda(t)$ and its velocity $\dot\lambda(t)$ are strongly oscillatory with time. In the simulations, we have used $w=-1/2$, $a=5.46~{\rm \AA}$, $L=80~\mu{\rm m}$, $\lambda_0=2.86~\mu{\rm m}$, $X_0=0$, $x_0=6~\mu{\rm m}$, $r_0=12.7~\mu{\rm m}$, $\omega_p=\pm\frac{\pi}{180}\times0.0195~\rm THz$, $\omega_0=4A/a^2=(2\pi)\times15.8~{\rm THz}$, $\omega_{\rm pd}=(2\pi)\times4.5\times10^{-7}~{\rm THz}$, $\alpha=1.9\times10^{-4}$, $K_X=(2\pi)^2\times7.2\times10^{-6}~\mu{\rm m}^{-1}{\rm THz}^2$, and $K_\lambda=(2\pi)^2\times3.0\times10^{-4}~\mu{\rm m}^{-1}{\rm THz}^2$.}
\label{fig:domainwall-motion}
\end{figure}

\textit{Example}---We now consider a concrete example based on a recent experiment where an AFM DW undergoes ultrafast motion comparable to the magnon velocity~\cite{Seyler2025high}. As sketched in Fig.~\ref{fig:domainwall-motion}(a), a quasi-1D AFM spin chain consists of an antiphase Néel-type DW, while an optical pump is applied at a distance from the DW center indicated by the dashed ellipse. The phenomenological model in Ref.~\cite{Seyler2025high} describes the DW by an 1D scaler field $\phi(x,t)$ and extracts the instantaneous DW position numerically from the field configuration without introducing CCs. This oversimplified treatment failed to tender a closed-form quantification for the collective behavior of the DW such as its center and width as functions of time, also missing the interplay among these variables via nonlinear coupling. In particular, the DW trajectories exhibit a pronounced helicity-dependent asymmetry that cannot be reproduced by the simple phenomenological picture.

To apply our formalism expressed in Eq.~\eqref{eq:ddn-final}, we first project the field dynamics onto two characteristic CCs $\xi^\mu(t)=[X(t),\lambda(t)]$ where $X$ is the DW center position and $\lambda$ is the DW width. In the experimental material Sr$_2$Cu$_3$O$_4$Cl$_2$, the Néel order is essentially confined in the easy plane due to a strong hard-axis anisotropy, so we can use $\bm n=(\cos\phi,\sin\phi,0)$ to simplify the problem. In addition, the pseudo-dipolar coupling gives rise to an effective in-plane anisotropy energy $U_{\rm pd}=\frac{1}{2}J_{\rm pd}M^2(n_x^2-n_y^2)$, which, in frequency units, gives rise to a generalized force $-\omega_{\rm pd}\omega_0M^2\bar{\Sigma}\bm n$ acting on $\bm{n}$ with $\omega_{\rm pd}=(\gamma/\mathcal{M}_0t_0)J_{\rm pd}$ and $\bar{\Sigma}\bm{n}=(n_x,-n_y,0)$. This term, together with the exchange force, ends up with
\begin{align}
\bm q_{\rm DW}=\frac14\omega_0AM^2\nabla^2\bm n-\omega_{\rm pd}\omega_0M^2\bar{\Sigma}\bm n
\label{eq:q-DW-main}
\end{align}
in the right-hand side of Eq.~\eqref{eq:n-eom-main}, hence a driving force $\mathcal{F}_{\rm DW}^\rho=(1/4L)\mathcal{G}^{\rho\beta}\int dx\bm{q}_{\rm DW}\cdot\partial_\beta\bm{n}$ in Eq.~\eqref{eq:ddn-final}.

Next, we invoke the DW profile ansatz based on the chosen CCs
\begin{align}
\phi(x,t)=2\arctan\exp\left\{\frac{-2w[x-X(t)]}{\lambda(t)}\right\}-w\pi,
\label{eq:DW-ansatz-main}
\end{align}
where $w=\mp1/2$ is the winding number of the DW~\cite{DWwidth}. The restoring force acting on the CCs can be introduced through a potential-energy penalty that drives the locally excited wall profile $\phi(x,t)$ back toward its equilibrium state $\phi_0(x)$, whose field-theoretical form takes $\mathcal{V}_{\rm trap}=(\kappa/8L)\int dx[\phi(x,t)-\phi_0(x)]^2$ where $L$ is the total length of the quasi-1D chain. Substituting the DW ansatz $\phi(x,t)=\phi[x;X(t),\lambda(t)]$ in Eq.~\eqref{eq:DW-ansatz-main} and expanding around the equilibrium $X=0$ and $\lambda=\lambda_0$ gives a harmonic potential~\cite{SM}
\begin{align}
\mathcal{V}_{\rm trap}=\frac{K_X}{8L}X^2+\frac{K_\lambda}{8L}(\lambda-\lambda_0)^2,\label{eq:trap-main}
\end{align}
where mixed terms like $X\lambda$ vanish due to parity. This potential leads to a driving force $\mathcal{F}^\rho_{\rm trap}=-\mathcal{G}^{\rho\beta}\partial_\beta \mathcal{V}_{\rm trap}$ in the right-hand side of Eq.~\eqref{eq:ddn-final}. Under the combined action of all generalized forces, the overall dynamics becomes
\begin{align}
\ddot{\xi}^\rho+\Gamma^\rho_{\mu\nu}\dot{\xi}^\mu\dot{\xi}^\nu+\alpha \omega_0M\dot{\xi}^\rho=\mathcal{F}_{\rm DW}^\rho+\mathcal{F}_{\rm trap}^\rho,
\label{eq:main-DW-CC-eom}
\end{align}
where $\rho,\mu,\nu$ run through the two CC variables $X$ and $\lambda$. The Berry curvature term vanishes identically because $\delta M=0$ in the compensated AFM limit.

The optical pump interacts with the DW for only about $0.1$ ps, much shorter than the time scale of the subsequent DW dynamics. Therefore, it is not included directly in Eq.~\eqref{eq:main-DW-CC-eom} as a driving force, but instead, sets a spatial profile of initial angular velocity $\dot{\phi}(x,0)$, while the initial displacement is negligible during this ultrashort pulse. In the experiment, the optical pump has a Gaussian profile centered at $x_0$ with a radius $r_0$, so the initial condition can be written as
\begin{align}
 \dot{\phi}(x,0)=\omega_p\exp\left[-\frac{2(x-x_0)^2}{r_0^2}\right],
 \label{eq:pump-main}
\end{align}
where $\text{sign}(\omega_p)=\pm1$ corresponds to the right/left-handed helicity of the driving photon. Now we need to determine the initial velocities of the CCs, $\dot X(0)$ and $\dot\lambda(0)$, from $\dot{\phi}(x,0)$. Mathematically, $\dot{\xi}^\mu(0)$ is determined by requiring the difference between the field-level velocity $\dot{\bm n}(0)$ and its projection onto the CC tangent space $\partial_\mu\bm n(0)\dot{\xi}^\mu(0)$ to be orthogonal to all tangent vectors $\partial_\beta\bm n(0)$,
\begin{align}
\left\langle\partial_\beta\bm{n}(0),\,\dot{\bm n}(0)-\partial_\mu\bm n(0)\dot{\xi}^\mu(0)\right\rangle=0,
\label{eq:main-proj-cond}
\end{align}
where $\langle \bm a,\bm b\rangle\equiv(1/L)\int dx\bm a(x)\cdot\bm b(x)$. After some simple algebra~\cite{SM}, we obtain
\begin{align}
\dot{\xi}^\mu(0)=\frac{1}{4L}\mathcal{G}^{\mu\nu}(0)\int dx \dot{\phi}(x,0)\partial_\nu\phi(x,0).
\label{eq:pump-projection-main}
\end{align}

By numerically solving Eq.~\eqref{eq:main-DW-CC-eom} under the initial condition Eq.~\eqref{eq:pump-projection-main}, we plot the DW motion in terms of $X(t)$ and $\lambda(t)$, and their time derivatives, in Fig.~\ref{fig:domainwall-motion}, where the full geodesic dynamics (G) is directly compared with the Newtonian approximation (N) with the connection term $\Gamma^\rho_{\mu\nu}\dot{\xi}^\mu\dot{\xi}^\nu$ discarded from the equation. As shown in Fig.~\ref{fig:domainwall-motion}(b) and (c), the Newtonian approximation yields symmetric trajectories when flipping the pump helicity $\omega_p\rightarrow-\omega_p$ [which changes the sign of $\dot{\xi}^\mu(0)$], namely, the dashed-blue and dashed-red curves are symmetric about the horizontal axis~\cite{symmetric}. As expected, we obtain the same symmetric pattern when flipping the DW winding number $w$ while keeping $\omega_p$ the same. By contrast, the connection term, being quadratic in the CC velocities, is even under the reversal of $w$ or $\omega_p$ at $t=0$, which leads to a pronounced asymmetry in the DW trajectories as reflected by the solid-blue and solid-red curves, consistent with the experimentally observed asymmetric DW displacement and velocity~\cite{Seyler2025high}. The apparent deviation of the geodesic dynamics from its Newtonian limit shows that for ultrafast AFM textures, the emerging gravity induced by the quantum metric is a non-negligible effect compared with the restoring and damping forces, which had been completed missed by previous studies.

Figures~\ref{fig:domainwall-motion}(d) and (e) reveal a highly oscillatory DW width $\lambda$ versus time (so is $\dot{\lambda}$) accompanying the translational motion. Since breathing DW width also occurs in the Newtonian limit, this is not by itself a unique signature of the emerging gravity of the quantum metric. Nevertheless, the appreciable oscillation amplitude of $\lambda(t)$ suggests that the DW width should never be approximated as a constant but must be treated as an independent CC. The predicted breathing of $\lambda$ is consistent with an unpublished experimental observation~\cite{PC}, substantiating an additional experimental probe into the dynamical behavior of ultrafast AFM textures.

\begin{acknowledgments}
The authors are indebted to K. L. Seyler for insightful discussions. This work is supported by the National Science Foundation under Award No. DMR-2339315.
\end{acknowledgments}


\begin{thebibliography}{47}
\bibitem [{cov()}]{covariance}%
  \BibitemOpen
  \href@noop {} {}\bibinfo {note} {{For an arbitrary non-singular
  reparameterization $\xi^\mu\to\xi^{\mu'}$, both $g_{\mu\nu}$ and
  $\ell_{\mu\nu}$ transform as tensors, i.e.,
  $T_{\mu'\nu'}=(\partial\xi^\mu/\partial\xi^{\mu'})(\partial\xi^\nu/\partial\xi^{\nu'})T_{\mu\nu}$.
  Also, the local metric, when nonsingular, is Riemannian rather than
  Lorentzian because $g_{\mu\nu}$ is by definition a Gram matrix, hence always
  being semi-definite: $v^\mu g_{\mu\nu}v^\nu=|v^\mu\partial_\mu\bm
  n|^2/4\geq0$ for any real $v^\mu$.}}\BibitemShut {Stop}%
\bibitem [{sym()}]{symmetric}%
  \BibitemOpen
  \href@noop {} {}\bibinfo {note} {{Numerically they are not perfectly
  symmetric because we have a finite system and the pump center is displaced
  from $X_0$.}}\BibitemShut {Stop}%
\end{thebibliography}
\end{document}